\documentclass[a4paper]{jpconf}
\usepackage{graphicx}
\begin{document}
\title{Anisotropic flow measurements in Pb--Pb collisions at 5.02 TeV with ALICE}

\author{You Zhou (for the ALICE Collaboration)}

\address{Niels Bohr Institute, University of Copenhagen, Blegdamsvej 17, 2100 Copenhagen, Denmark}

\ead{you.zhou@cern.ch}

\begin{abstract}

Anisotropic flow is a sensitive probe of the initial conditions and the transport properties of the Quark Gluon Plasma (QGP) produced in heavy-ion collisions. In these proceedings, we present the first results of elliptic ($v_2$), triangular ($v_3$) and quadrangular flow ($v_4$) of charged particles in Pb--Pb collisions at $\sqrt{s_{_{\rm NN}}}=$ 5.02 TeV with the ALICE detector. In addition, the comparison of experimental measurements to various theoretical calculations will be discussed. This provides a unique opportunity to test the validity of the hydrodynamic picture and discriminates between various possibilities for the temperature dependence of shear viscosity to entropy density ratio of the produced QGP.

\end{abstract}

\section{Introduction}

Collisions of high-energy heavy-ions, at the Brookhaven Relativistic Heavy Ion Collider (RHIC) 
and the CERN Large Hadron Collider (LHC), allow us to create and study the properties of Quark Gluon Plasma (QGP) in the laboratory.
The azimuthal anisotropy in particle production is, at these energies, an observable 
which provides experimental information on the Equation of State and the transport properties of the QGP.
This anisotropy is usually characterized by the Fourier flow-coefficients~\cite{Voloshin:1994mz},
\begin{eqnarray}
v_{n} &=& \langle\cos[n(\varphi-\Psi_{n})\rangle, 
\;\;\; {\rm or\ equivalently} \nonumber \\ 
v_n  &=& \langle e^{in\varphi}  e^{-in \Psi_n} \rangle,
\label{eq1}
\end{eqnarray}
where $\varphi$ is the azimuthal angle of the particles, $\Psi_{n}$ is the $n^{\rm{th}}$-order flow plane (or named final state symmetry plane) angle
and $\langle~\rangle$ denotes an average over the selected particles and events. 
Experimental measurements together with comparisons to theoretical calculations show that anisotropic flow sheds new light on the initial conditions and is sensitive to the shear viscosity to entropy density ratio $\eta/s$ of the QGP. Recent anisotropic flow measurements~\cite{ALICE:2016kpq, Adam:2016ows, Adam:2016nfo} in Pb--Pb collisions at $\sqrt{s_{\rm NN}} =$ 2.76 TeV provide additional information to improve the theoretical descriptions of data.
In these proceedings, we present the latest anisotropic flow measurements in Pb--Pb collisions at $\sqrt{s_{\rm NN}} =$ 5.02 TeV in ALICE~\cite{Adam:2016izf}. The comparisons to hydrodynamic and AMPT model calculations will be discussed below.

\section{Analysis Details}

The data used in this analysis were recorded with the ALICE detector in November 2015 in run 2 at the LHC with Pb-Pb collisions at $\sqrt{s_{\rm NN}}= 5.02$ TeV. About 140 k Pb--Pb events were recorded with a minimum-bias trigger, based on signals from two VZERO detectors (-3.7$\textless \eta \textless$-1.7 for VZERO-C and 2.8$\textless \eta \textless$5.1 for VZERO-A) and on the Silicon Pixel Detector. The VZERO detectors were also used for the determination of the collision centrality in Pb--Pb collisions. Charged particles are reconstructed using the Inner Tracking System and the Time Projection Chamber with full azimuthal coverage for pseudo-rapidity range $|\eta|\textless$0.8.

\section{Results and Discussions}

\begin{figure}[th]
\begin{center}
\includegraphics[width=0.5\textwidth]{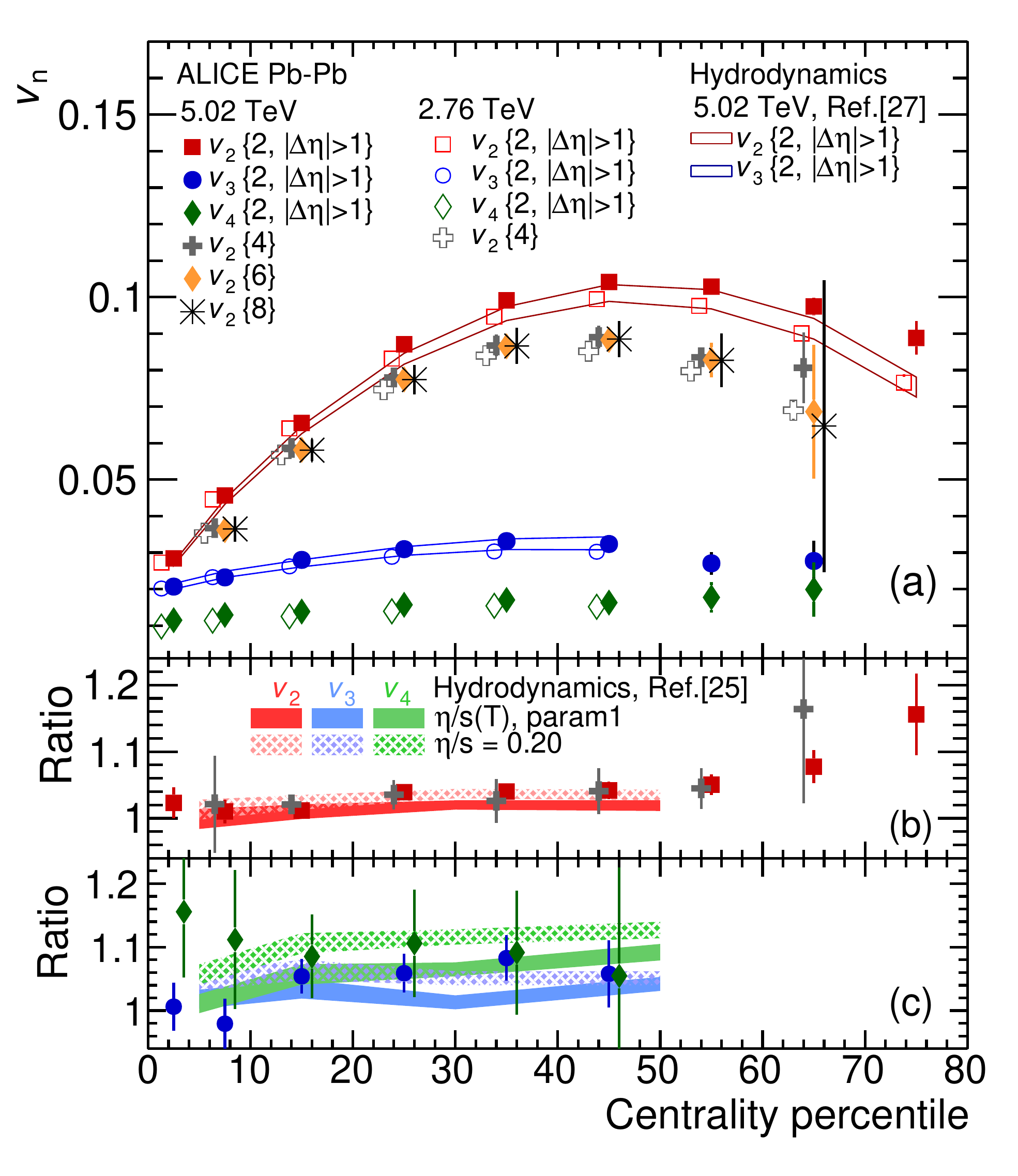}
\caption{(color online) (a) Anisotropic flow $v_{n}$ integrated over the $p_{\rm T}$ range 0.2 $< p_{\rm T} <$ 5.0 GeV/$c$, as a function of event centrality, for the two--particle (with $|\Delta\eta|>1$) and multi-particle cumulant methods. The ratios of $v_n$ from Pb--Pb collisions at 5.02 TeV and 2.76 TeV, are presented in Fig.~\ref{fig1} (b) and (c). Various hydrodynamic calculations are also presented \cite{Noronha-Hostler:2015uye,Niemi:2015voa}.}
\label{fig1} 
\end{center}
\end{figure}

The centrality dependence of $v_{2}$, $v_{3}$ and $v_{4}$ are presented in Fig.~\ref{fig1}. Both two-particle correlations with pseudorapidity and multi-particle cumulants methods are used in these measurements. 
It is seen that $v_{2}\{2, |\Delta\eta|>1\}$ increases from central to peripheral collisions, saturates in the 40--50\% centrality class with a maximum value of 0.104 $\pm$ 0.001 (stat.) $\pm$ 0.002 (syst.).
Similar centrality dependence is observed in the measurements of multi-particle cumulants $v_{2}\{4\}$, $v_{2}\{6\}$ and $v_{2}\{8\}$ and results are agree within 1\%.
Assuming that non-flow effects are suppressed by the pseudorapidity gap, the remaining differences between two- and multi-particle cumulants of $v_{2}$ can be attributed to the different sensitivities to elliptic flow fluctuations. 
Compared to $v_2$, higher harmonics i.e.\ $v_3$ and $v_4$ are smaller and the centrality dependence is much weaker.
In addition, the hydrodynamic calculations~\cite{Noronha-Hostler:2015uye} which combine the changes in initial spatial anisotropy and the hydrodynamic response are presented in Fig.~\ref{fig1}.\ (a) for comparison. The calculations are found to be compatible with the measured anisotropic flow $v_n$ coefficients. Furthermore, using a new set of Lund string fragmentation parameters and partonic cross section 3 mb, the AMPT model with string melting scenario can quantitatively reproduce the anisotropic flow measurements~\cite{Feng:2016emh}.

To better illustrate the energy evolution of anisotropic flow, Fig~\ref{fig1}.\ (b) and (c) shows the ratios of $v_n$ measured at 5.02 TeV to 2.76 TeV.
It is observed that the increase of $v_{2}$ and $v_{3}$ from the two energies is rather moderate, while for $v_{4}$ it is more pronounced. 
Considering none of the ratios of flow harmonics from two energies exhibit a significant centrality dependence in the centrality range 0--50\%, a fit with a constant value are performed for these ratios. An increase of (3.0$\pm$0.6)\%, (4.3$\pm$1.4)\% and (10.2$\pm$3.8)\% over the centrality range 0--50\% in Pb--Pb collisions is obtained for $v_{2}$, $v_{3}$ and $v_{4}$, respectively.
This increase of anisotropic flow is in agreement with theoretical predictions in Ref~\cite{Niemi:2015voa}. The data seems to support a low value of $\eta/s$ for the QGP created in Pb--Pb collisions at $\sqrt{s_{_{\rm NN}}} = $ 5.02~TeV and indicates that $\eta/s$ does not change significantly with respect to Pb--Pb collisions at $\sqrt{s_{_{\rm NN}}}=2.76$~TeV.

\begin{figure}[th]
\begin{center}
\includegraphics[width=0.47\textwidth]{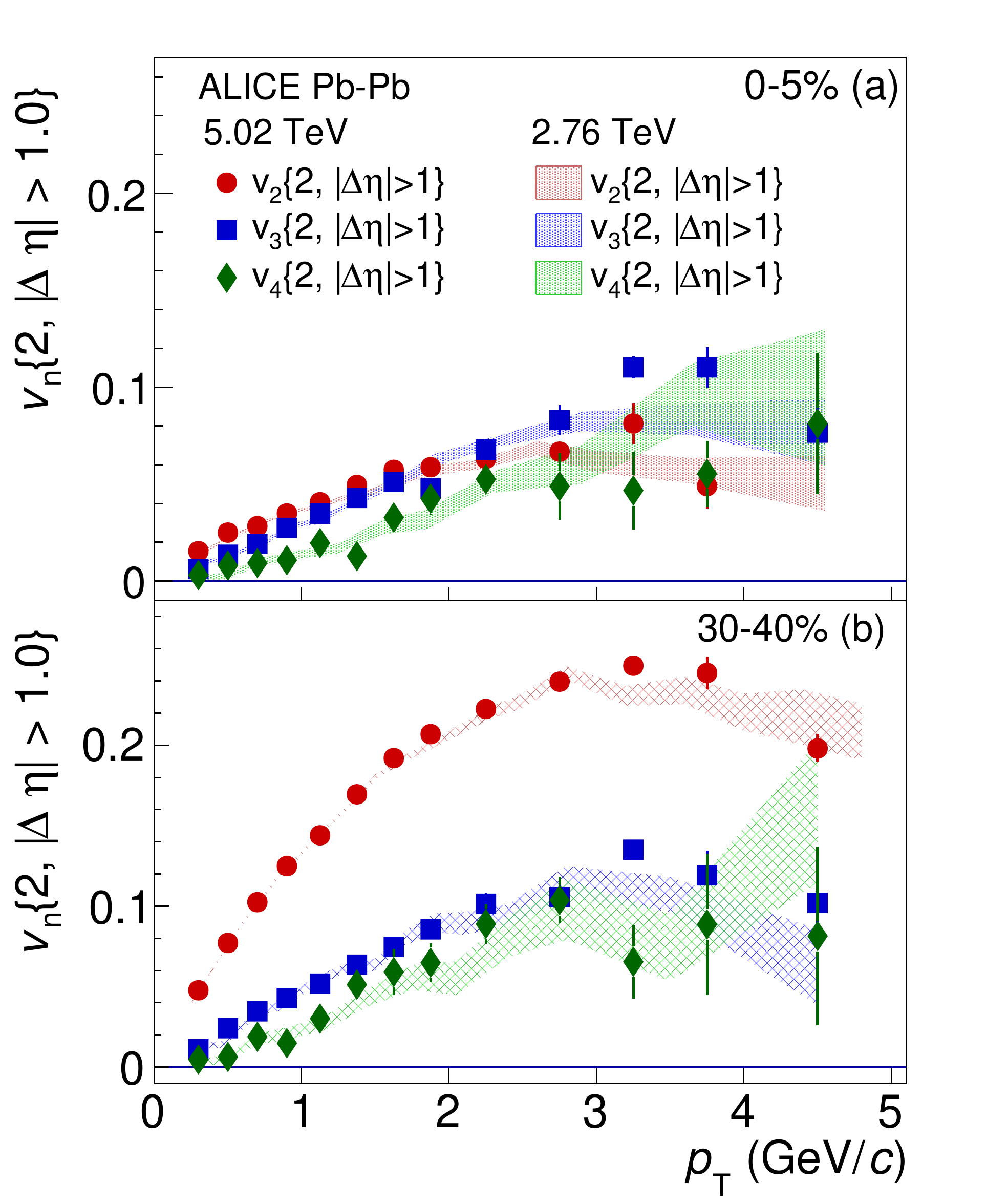}
\includegraphics[width=0.47\textwidth]{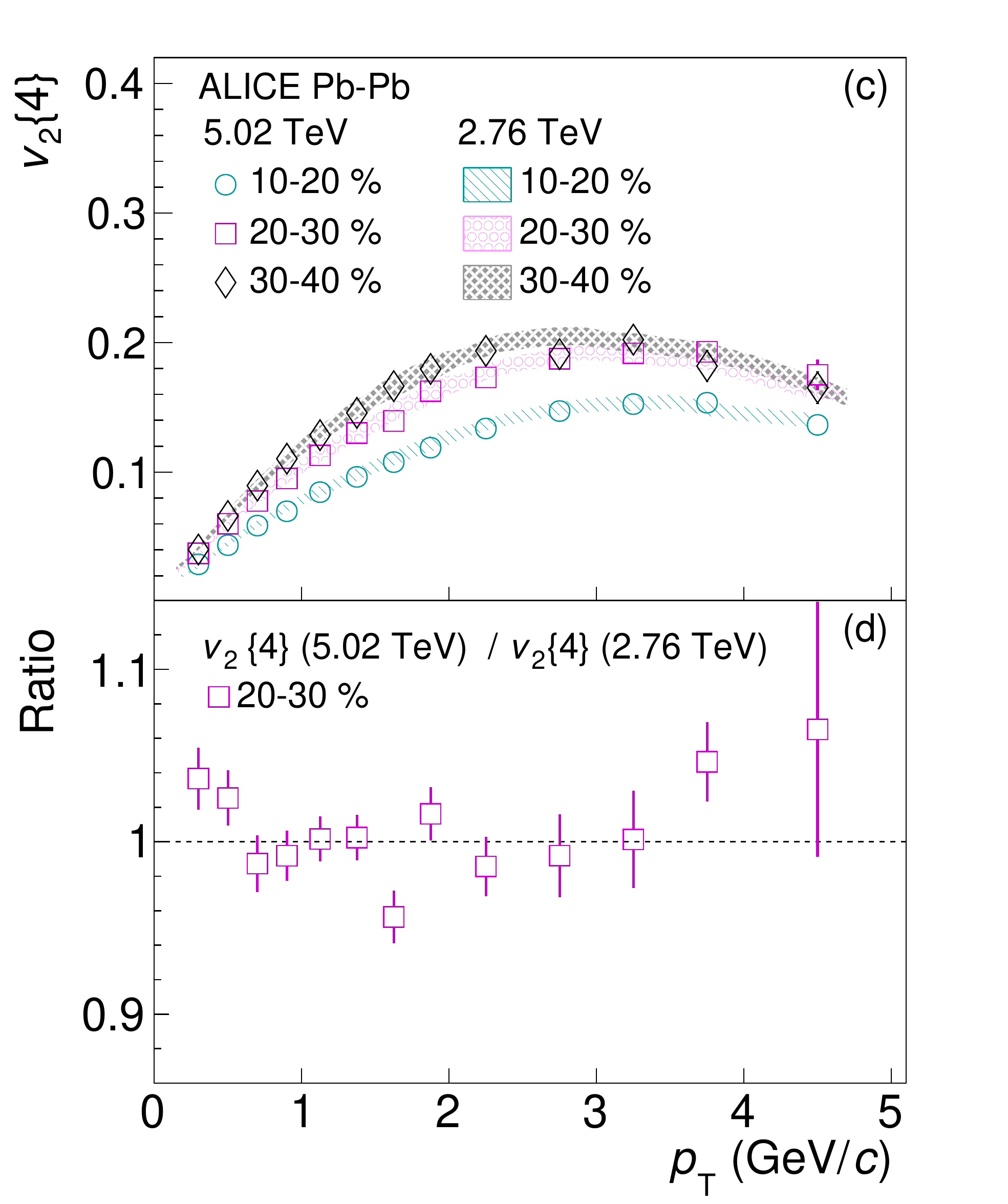}
\caption{(color online) $v_{n}(p_{\rm T})$ for the centrality 10--20\%, 20--30\% and 30--40\%. Measurements for Pb--Pb collisions at $\sqrt{s_{_{\rm NN}}}=$ 2.76 TeV are presented as shading. (d) The ratio of $v_{2}\{4\}$ in 20--30\% from two collision energies is also shown here.}.
\label{fig2} 
\end{center}
\end{figure}

The transverse momentum ($p_{\rm T}$) dependent $v_{2}\{2, |\Delta\eta|>1\}$, $v_{3}\{2, |\Delta\eta|>1\}$ and $v_{4}\{2, |\Delta\eta|>1\}$ are presented for the 0--5\% and 30--40\% centrality classes. 
In 0--5\% centrality class, $v_{3}\{2\}$ is larger than $v_{2}\{2\}$ for $p_{\rm T} >$ 2 GeV/$c$, and $v_{4}\{2\}$ is roughly compatible with $v_{2}\{2\}$. For the 30--40\% centrality class, $v_{2}\{2\}$ is higher than $v_{3}\{2\}$ and $v_{4}\{2\}$ for the entire $p_{\rm T}$ range measured, with no crossing of the different order flow coefficients. 
The $p_{\rm T}$-differential $v_{2}\{4\}$ decreases from mid-central to central collisions over the $p_{\rm T}$ range measured.
The results are consistent with the previous measurements from 2.76 TeV, as also illustrated by the ratio of $v_{2}\{4\}$ for the two energies in Fig.~\ref{fig2} (d). 
This suggests that the increase observed in the $p_{\rm T}$ integrated flow results seen in Fig.~\ref{fig1} could be due to stronger radial flow produced at 5.02 TeV. It leads to an increase of mean transverse momentum $\langle p_{\rm T} \rangle$ from 2.76 to 5.02 TeV. 
The measurements of $p_{\rm T}$-differential flow are expected to be more sensitive to initial conditions and $\eta/s$ of the QGP. The results are found to be roughly compatible with AMPT calculations, meanwhile no hydrodynamic calculations are available yet.
The future comparisons of $p_{\rm T}$-differential flow between experimental measurements and hydrodynamic calculations will provide important information to constrain further details of the theoretical calculations, e.g. determination of radial flow and freeze-out conditions.

\begin{figure}[th]
\begin{center}
\includegraphics[width=0.6\textwidth]{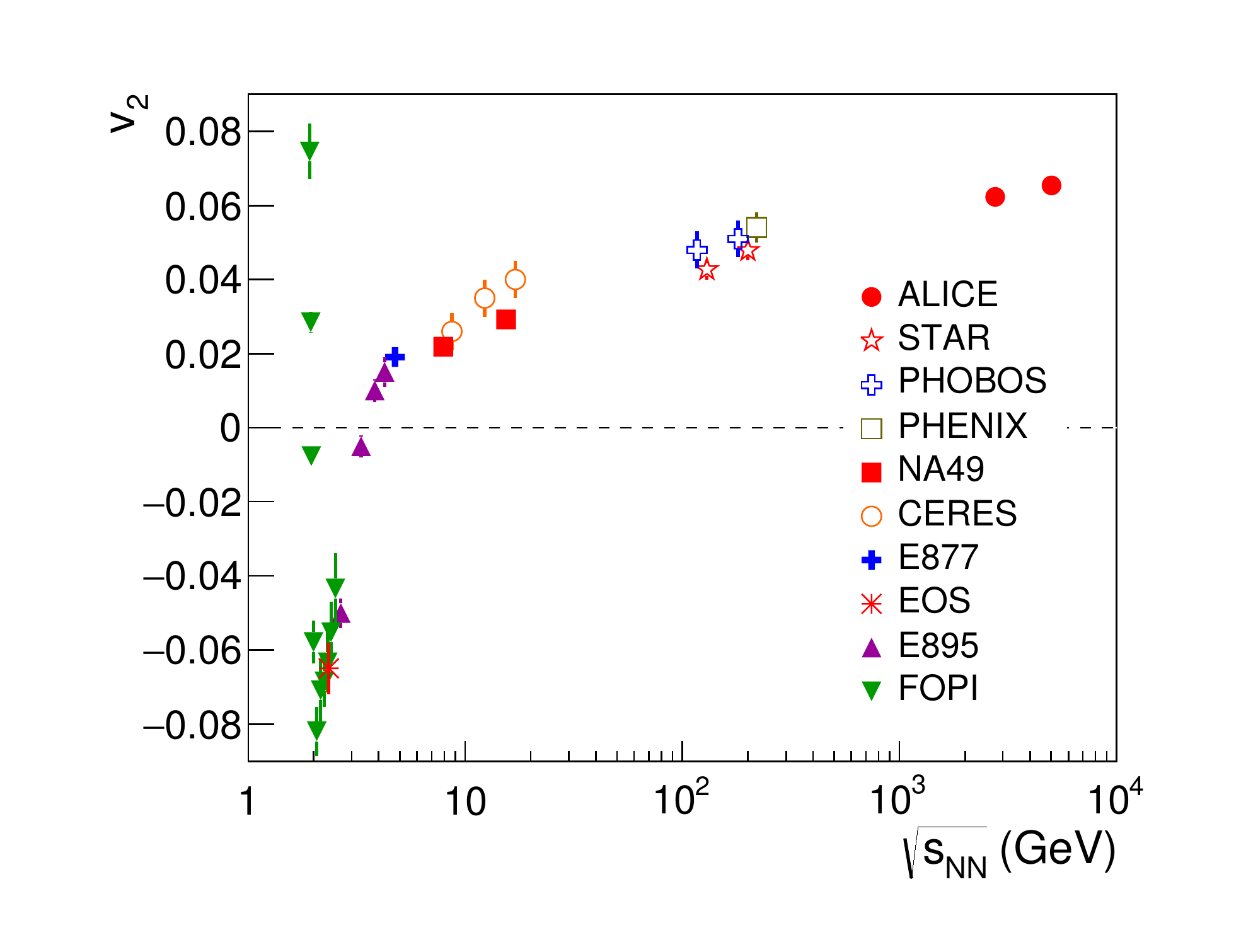}
\caption{(color online) $p_{\rm T}$ integrated $v_2$ in centrality 20-30\% as a function of collisions energy. }
\label{fig3} 
\end{center}
\end{figure}

Figure~\ref{fig3} presents the collisions-energy dependence of the fully $p_{\rm T}$-integrated $v_2$ measured in 20--30\% centrality with different collision systems and energies. 
A continuous increase of anisotropic flow for this centrality has been observed from SPS/RHIC to LHC energies. For these fully $p_{\rm T}$ integrated coefficients, an increase of $4.9\pm1.9\%$ is observed going from $\sqrt{s_{_{\rm NN}}}=2.76$ to 5.02 TeV, which is close to values of the previously-mentioned hydrodynamic calculations \cite{Noronha-Hostler:2015uye,Niemi:2015voa}.

\section{Summary}

In summary, the first anisotropic flow measurements of charged particles in Pb--Pb collisions at $\sqrt{s_{_{\rm NN}}} =$  5.02 TeV at the LHC are presented in this proceedings. An average increase of (3.0$\pm$0.6)\%, (4.3$\pm$1.4)\% and (10.2$\pm$3.8)\%, is observed for the transverse momentum integrated elliptic, triangular and quadrangular flow, respectively, over the centrality range 0--50\% going from 2.76 TeV to 5.02 TeV. The $p_{\rm T}$-differential flow has also been investigated, it does not change appreciably between the two LHC energies. Therefore, the increase in integrated flow coefficients could be mainly due to an increase in average $p_{\rm T}$. This agrees with predictions from hydrodynamic models~\cite{Noronha-Hostler:2015uye,Niemi:2015voa}. Further comparisons of $p_{\rm T}$-differential flow measurements and theoretical calculations will provide extra constraints on the initial conditions and the transport properties of the QGP.

\end{document}